\documentstyle[twocolumn,prb,aps,epsfig,amsfonts,multicol,graphicx]{revtex}
\begin{document}
\draft
\preprint{}

\newcommand{\1}{{\bf \scriptstyle 1}\!\!{1}}
\newcommand{\p}{\partial}
\newcommand{\D}{^{\dagger}}
\newcommand{\bx}{{\bf x}}
\newcommand{\bk}{{\bf k}}
\newcommand{\bn}{{\bf n}}
\newcommand{\bv}{{\bf v}}
\newcommand{\bw}{{\bf w}}
\newcommand{\bp}{{\bf p}}
\newcommand{\bq}{{\bf q}}
\newcommand{\bu}{{\bf u}}
\newcommand{\bA}{{\bf A}}
\newcommand{\bB}{{\bf B}}
\newcommand{\bD}{{\bf D}}
\newcommand{\bE}{{\bf E}}
\newcommand{\bK}{{\bf K}}
\newcommand{\bL}{{\bf L}}
\newcommand{\bP}{{\bf P}}
\newcommand{\bQ}{{\bf Q}}
\newcommand{\bS}{{\bf S}}
\newcommand{\bH}{{\bf H}}
\newcommand{\bg}{{\bf g}}
\newcommand{\balpha}{\mbox{\boldmath $\alpha$}}
\newcommand{\bsigma}{\mbox{\boldmath $\sigma$}}
\newcommand{\bSigma}{\mbox{\boldmath $\Sigma$}}
\newcommand{\bOmega}{\mbox{\boldmath $\Omega$}}
\newcommand{\bpi}{\mbox{\boldmath $\pi$}}
\newcommand{\bphi}{\mbox{\boldmath $\phi$}}
\newcommand{\bnabla}{\mbox{\boldmath $\nabla$}}
\newcommand{\bmu}{\mbox{\boldmath $\mu$}}
\newcommand{\bepsilon}{\mbox{\boldmath $\epsilon$}}

\newcommand{\iLambda}{{\it \Lambda}}
\newcommand{\cA}{{\cal A}}
\newcommand{\cD}{{\cal D}}
\newcommand{\cL}{{\cal L}}
\newcommand{\cH}{{\cal H}}
\newcommand{\cI}{{\cal I}}
\newcommand{\cM}{{\cal M}}
\newcommand{\cO}{{\cal O}}
\newcommand{\cR}{{\cal R}}
\newcommand{\cU}{{\cal U}}
\newcommand{\cT}{{\cal T}}

\newcommand{\be}{\begin{equation}}
\newcommand{\ee}{\end{equation}}
\newcommand{\bea}{\begin{eqnarray}}
\newcommand{\eea}{\end{eqnarray}}
\newcommand{\beqa}{\begin{eqnarray*}}
\newcommand{\eeqa}{\end{eqnarray*}}
\newcommand{\nn}{\nonumber}
\newcommand{\DD}{\displaystyle}

\newcommand{\ba}{\left[\begin{array}{c}}
\newcommand{\baa}{\left[\begin{array}{cc}}
\newcommand{\baaa}{\left[\begin{array}{ccc}}
\newcommand{\baaaa}{\left[\begin{array}{cccc}}
\newcommand{\ea}{\end{array}\right]}

\twocolumn[
\hsize\textwidth\columnwidth\hsize\csname
@twocolumnfalse\endcsname

\title{Discrete Fourier Transform in Nanostructures using Scattering}

\author{Michael N.~Leuenberger$^{(1,2)}$, 
Daniel Loss$^{(2)}$, 
Michael E. Flatt\'e$^{(1)}$, 
and D.~D.~Awschalom$^{(3)}$
}
\address{$^{(1)}$ Department of Physics and Astronomy, University of Iowa,\\
Iowa City, IA 52242, USA}
\address{$^{(2)}$ Department of Physics and Astronomy, University of Basel \\
Klingelbergstrasse 82, 4056 Basel, Switzerland}
\address{$^{(3)}$ Department of Physics, University of California, Santa Barbara, 
CA 93106-9530, USA}

\date{\today}
\maketitle

\begin{abstract}
In this paper we show that the discrete Fourier transform
can be performed by scattering a coherent particle or laser beam
off a two-dimensional potential that has the shape of rings or peaks.
After encoding the initial vector into the two-dimensional potential,
the Fourier-transformed vector can be read out by detectors
surrounding the potential.
The wavelength of the laser beam determines the necessary accuracy of
the 2D potential, which makes our method very fault-tolerant.
\end{abstract}

 ] \narrowtext 

\section{Introduction}

The fast developing fields of quantum computing and quantum 
information processing have attracted much interest recently
\cite{Nielsen,Bouwmeester,Braunstein,Williams,Berman,Lo}.
Most of the papers in the field of quantum computing, such as Shor's and Grover's 
about factoring numbers\cite{Shor} and searching a database\cite{Grover1}, 
use many-particle physics (i.e. qubits) for their implementation. 
However, it is also possible to implement quantum information processes in the 
so-called unary representation of dimension $N$, which consists of a statistical ensemble
of non-interacting $N$-level systems that can be described by a single-particle Hamiltonian.
So only interference, but no entanglement is needed for the implementation
of quantum algorithms in the unary representation.
It was already shown that the Grover algorithm\cite{Grover2} can 
be performed in a single shot using the unary representation.
While in Ref.~\onlinecite{Ahn} the quantum database search has been done experimentally
with atoms in a beam that serve as non-interacting
$N$-level systems, Refs.~\onlinecite{Leuenberger2000} and \onlinecite{Leuenberger2001}
show theoretically that the large spin of molecular magnets and nuclear spins in semiconductors
can be used as non-interacting $N$-level systems,
respectively. 
Here we propose a method to perform the classical
discrete Fourier transform (DFT)
in the time ensemble of quantum or classical
scattering events, where
the scattering potential provides the classical $N$-level system
for the read-in of information and the $N$ detectors
provide the classical $N$-level system for the read-out of the information.
In the case of quantum scattering a single-particle Hamiltonian describes the 
information processing.
In contrast to Shor's algorithm
our approach requires an exponential
overhead of the hardware. However, once our device is built the DFT can
be computed very efficiently, i.e. in a single shot.
In our method we propose to scatter a coherent matter or laser beam 
off a two-dimensional potential consisting of $N$ concentric rings or $N$ peaks, 
the amplitudes of which 
represent the elements of a vector in an $N$-dimensional vector space
where the information is encoded.
The Fourier transformation of 
this vector can then be read out immediately by detectors surrounding 
the 2D potential. 
Our method is closely related to the general field of Fourier optics\cite{Gu,Goodman,Gaskill},
where a diffraction pattern is Fourier transformed into a desired image pattern
by means of laser scattering.
We go a step further in the sense that we describe how to use
a 2D scattering potential to perform the DFT efficiently.
The concentric rings and peaks can be produced by charge
or spin distributions. 
The calculation time $\tau_{c}$ is roughly 
given by the diameter of the detector ring (e.g. 1 mm) divided 
by the light velocity $c=3\times 10^8$ m/s, i.e. $\tau_{c}\sim 10^{-11}$ s. 
Using for example a laser with a wavelength 
of 500 nm and a spot radius of 100 $\mu$m enables the Fourier 
transformation of a function with 100 sampling points. The larger 
the spot size and the smaller the wavelength of the laser, the 
more sampling points can be achieved. The advantage of this device 
would be that it can be used already for example as DFT coprocessor 
to speed up present computers operating at room temperature.
Rough estimates show a possible speedup of $10^6$  or more (see below).

In Sec.~\ref{section_DFT} we give an overview on the DFT and the
Fast Fourier Transform (FFT). Sec.~\ref{cms} and Sec.~\ref{lbs}
describe the quantum-mechanical coherent matter scattering
and the classical laser beam scattering off a 
specially designed 2D potential that is used for the
implementation of the DFT.

\section{Discrete Fourier Transformation}
\label{section_DFT}

The discrete Fourier transformation (DFT) has a wide range of 
applications in the field of digital signal processing, such 
as spectral analysis and filtering. 
The DFT enables the possibility 
to extract the period of a function, which can be used to factor 
numbers. Thus finding efficient ways to compute the DFT 
is the key to the efficient factorization of numbers.
The calculation time of a DFT can 
be greatly reduced by the classical algorithm called Fast Fourier 
Transform (FFT), which gets rid of the redundancies found in 
the DFT. In a similar way Shor's quantum algorithm reduces the 
calculation time of the discrete quantum Fourier transformation 
(DQFT) by means of an efficient quantum circuit\cite{Shor} that uses 
qubits as information carriers.
In our proposal the superposition principle of quantum or classical
wavefunctions is used to perform an $N$-dimensional DFT 
with $N$ shots or even a single shot in one scattering
device. The FFT provides then the possiblity to link
several scattering devices in parallel.
This means that the speedup factor $N$ can be multiplied by the number
of the parallel scattering devices.

First it is instructive to give the definition of a DFT. 
When we calculate the DFT of a function $f(z)$ 
that maps complex numbers on complex numbers, we can first discretize 
the function $f(z)$ in $M$ total sampling points $z_{1}, z_{2}, \dots, z_M$, 
leading to $M$ values $f_{1}=f(z_{1}), f_{2}=f(z_{2}), \dots , f_M=f(z_M)$. Then 
the DFT of $f$ is given by
\be
F_{k} =\sum_{j=0}^{M-1}e^{2\pi ijk/M} f_{j}.
\ee
The matrix elements of a DFT consist of the irreducible representation 
of the cyclic group produced by a rotation R$_{2\pi/M}$ 
about the angle $2\pi/M$, where $M$ is the dimension 
of the vector space, which is equal to the number of sampling 
points. For example a DFT of dimension $M=4$ is represented 
by 
\be
\left[ 
\begin{array}{cccc}
1 & 1 & 1 & 1 \\
1 & e^{\frac{1}{4}2\pi i} & e^{\frac{1}{2}2\pi i} & e^{\frac{3}{4}2\pi i} \\
1 & e^{\frac{1}{2}2\pi i} & e^{2\pi i} & e^{\frac{1}{2}2\pi i} \\
1 & e^{\frac{3}{4}2\pi i} & e^{\frac{1}{2}2\pi i} & e^{\frac{1}{4}2\pi i}
\end{array}
\right]
\ee
It will turn out that the exponential overhead of the hardware imposes a physical limit 
to the number of sampling points $N$ that can be processed 
by our DFT device. However, the {\it Danielson-Lanczos Lemma}\cite{Press}, 
which is widely used for the FFT algorithm, states that the DFT 
for $M$ sampling points can be separated into the sum over the 
DFT for the $M/2$ even sampling points and the DFT for the $M/2$ 
odd sampling points, i.e.
\be
F_{k} =\sum\limits_{j=0}^{M/2-1}e^{2\pi ik(2j)/M} f_{2j} 
+\sum\limits_{j=0}^{M/2-1}e^{2\pi ik(2j+1)/M} f_{2j+1}.
\ee
Usually, the FFT uses this {\it Danielson-Lanczos Lemma} recursively 
down to Fourier transforms of length 2$^{0}$. Therefore it is essential 
that $M=2^m$, where $m$ is a positive integer. 
As we want to use 2D scattering potentials with real amplitudes,
we can separate the real and imaginary part of the DFT:
\be
F_{k} =\sum_{j=0}^{M-1}e^{2\pi ijk/M} g_{j}+i\sum_{j=0}^{M-1}e^{2\pi ijk/M} h_{j},
\ee
where $f(z_j)=g(z_j)+ih(z_j)$ with real functions $g(z_j)$ and $h(z_j)$.

Since 
our device can perform the DFT efficiently for $N$ sampling 
points, we can stop the recursion at $N=2^n$ sampling points, 
where $n$ is a positive integer that is smaller than $m$. 
Using our device as coprocessor of an existing computer, the 
remaining work of the main processor consists in combining the $M/N$ 
partial Fourier transforms, which takes of order $\log_{2}(M/N)M/N=(m-n)M/N$ operations. 
Thus, compared with initially 
necessary $M\log_{2}(M)$ operations a speedup of at least 
a factor $N$ is achieved. In this estimate of the speedup, 
the necessary time to calculate the DFT for the $N$ sampling 
points is assumed to be of the order of 10 ps, i.e. negligible. 
Alternatively, one could say that an existing computer equipped 
with our device is capable of factoring numbers that are $N$ 
times larger than the largest nowadays factorizable number.
If $M/N$ DFT devices are linked in parallel, a speedup 
of factor $M$ is achieved.

\begin{figure}[htb]
  \begin{center}
    \leavevmode
\epsfxsize=8.5cm \epsffile{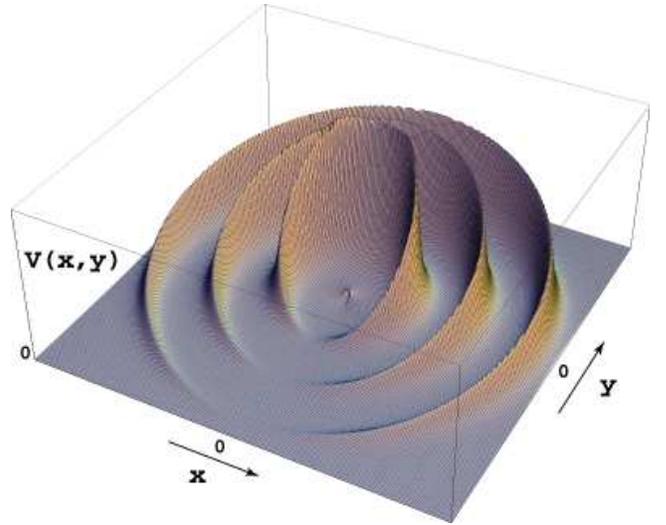}
  \end{center}
\caption{2D scattering potential $V(\bx')=(1+\sin\alpha)\sum_{j=1}^Na_j\delta(r'-r_j)/\sqrt{r'}$, which performs the DFT in N shots in the case of elastic scattering and in a single shot in the case of inelastic scattering.} 
\label{fig1}
\end{figure}

\section{Coherent matter scattering}
\label{cms}

Now we describe the various possible implementations of our DFT 
device, in which a coherent matter or laser beam is scattered 
off a 2D potential. This potential consists of radial $\delta$-functions
or $\delta$-peaks on a straight line, 
the amplitudes of which both sample the points of the
real or imaginary part of our function $f(z)$
and represent the amplitudes of our basis vectors. 
Let $\bv=\sum_{j=1}^Na_j\bw_j$ be the input vector, where
$\bw_j$ are the basis vectors with amplitudes $a_j=g(z_j)$ or $a_j=h(z_j)$.
Then our 2D potential is of the form 
$V(\bx)\propto\sum_{j=1}^Na_j\delta(\bx-\bx_j)$.
The read-out is done by detectors surrounding this potential. 
While the coherent matter beam needs a high precision of 
the potential, the laser beam provides a high fault-tolerance
of the potential since faults that are smaller than the 
wavelength of the laser beam remain undetected. We give the derivation
for both the coherent matter and the laser beam scattering.
We first start with the coherent matter beam. 
We use the nomenclature of Ref. \onlinecite{Sakurai}. First we calculate the 2D 
scattering amplitude, from which we can infer the precise shape 
of the 2D potential required for the implementation of the DFT. 
For this we have to start from the well-known Lippman-Schwinger 
equation\cite{Sakurai}
\be
\left\langle \bx\left| \psi ^{(\pm )}\right. \right\rangle =\left\langle
\bx\right.\left| \phi \right\rangle +\frac{2m}{\hbar ^{2} } \int d^{2}
x'G_{\pm } (\bx,\bx')\left\langle \bx'\right| V\left| \psi ^{(\pm )}
\right\rangle,
\label{Lippman-Schwinger}
\ee
where $\left| \phi \right\rangle $
is the incoming wave, $V$ is the 2D scattering potential, 
and 
\be
G_{\pm } (\bx,\bx')=\frac{\hbar ^{2} }{2m} \left\langle \bx\right|
\frac{1}{E-H_{0} \pm i\epsilon } \left| \bx'\right\rangle
\ee
is the Green's function, which also solves the Helmholtz equation 
$(\bnabla ^{2} +k^{2} )G_{\pm } (\bx,\bx')=\delta ^{(2)} (\bx-\bx')$.
For the evaluation of the Green's function we transform first
to the momentum representation, i.e.
\be
G_{\pm } (\bx,\bx')
=\frac{\hbar^{2}}{2m}\int\frac{d^2p'}{(2\pi\hbar)^2}
\frac{e^{\frac{i}{\hbar}\bp'\cdot(\bx-\bx')}}{E-\frac{p'^2}{2m}\pm i\epsilon},
\ee
where we have used the 2D overlap $\left<\bx|\bp\right>=e^{i\bp\cdot\bx/\hbar}/2\pi\hbar$.
Inserting $E=\hbar^2k^2/2m$, $\bp'=\hbar\bQ$, and going
to polar coordinates $(Q,\phi)$, we get
\bea
G_{\pm } (\bx,\bx') & = & \frac{1}{(2\pi)^2} \int_0^\infty QdQ
\int_0^{2\pi}d\phi \frac{e^{iQ|\bx-\bx'|\cos\phi}}{k^2-Q^2\pm i\epsilon}\nn\\
& \approx & \frac{1}{(2\pi)^2} \int_0^\infty QdQ
\frac{2^{3/2}\sqrt{\pi}}{\sqrt{Q|\bx-\bx'|}} \nn\\
& & \times\frac{\cos(Q|\bx-\bx'|-\frac{\pi}{4})}{k^2-Q^2\pm i\epsilon} \nn\\
& = & -\frac{1}{4\sqrt{\pi } } \frac{e^{\pm i(k\left| \bx-\bx'\right| -\frac{\pi
}{4} )} }{\sqrt{2k\left| \bx-\bx'\right| } }.
\eea 
For the approximation we have used the integral representation of the Bessel 
function 
$J_{0} (Q\left| \bx-\bx'\right| )=\frac{1}{2\pi } \int\nolimits_{0}^{2\pi
}d\phi \;e^{iQ\left| \bx-\bx'\right| \cos \phi }  \approx
\sqrt{\frac{2}{\pi Q\left| \bx-\bx'\right| } } \cos (Q\left| \bx-\bx'\right|
-\frac{\pi }{4} )$, which holds in the asymptotic regime 
$Q\left| \bx-\bx'\right| >>1$, i.e. 
the local 2D potential 
$\left\langle \bx'\right| V\left| \bx''\right\rangle =V(\bx')\delta ^{(2)}
(\bx'-\bx'')$
and the detectors at $\bx$ are far apart from each other.
Then we obtain with Eq.~(\ref{Lippman-Schwinger})
\be
\left\langle \bx\left| \psi ^{(+)} \right.\right\rangle =\frac{1}{2\pi }
\left[ e^{i\bk\cdot \bx} +\frac{e^{i(kr-\frac{\pi }{4} )} }{\sqrt{kr} }
f(\bk',\bk)\right],
\label{scattered_wavefunction}
\ee
where $r:=\left| \bx\right| $, 
$r':=\left| \bx'\right| $, 
$\bk':=k\frac{\bx}{\left| \bx\right| } $, and
\be
f(\bk',\bk)=-\frac{\sqrt{2\pi } }{4} \frac{2m}{\hbar ^{2} } \int d^{2}
x'e^{-i\bk'\cdot \bx'} V(\bx')\left\langle \bx'\left| \psi ^{(+)}
\right.\right\rangle
\label{scattering_amplitude}
\ee
is the 2D scattering amplitude. 
The amplitude of the 2D spherical wave decays as $1/\sqrt{r}$.
This is in agreement with the topological argument that
the circumference of a circle is $2\pi r$ and that
the probability has to be conserved on the circle.
In first-order Born approximation,
where $|f(\bk',\bk)|\ll\sqrt{kr}$,
we can insert 
$\left\langle \bx'\right.\left| \psi ^{(+)} \right\rangle =\frac{e^{i\bk\cdot
\bx'} }{2\pi } $
into Eq. (\ref{scattering_amplitude}), 
and we define $\bq:=\bk'-\bk$ as the scattering wavevector.
Eq.~(\ref{scattering_amplitude}) becomes then
\be
f^{(1)} (\bk',\bk)= 
-\frac{1}{4\sqrt{2\pi } } \frac{2m}{\hbar ^{2} } \int
d^{2} x'e^{iqx'\cos\varphi} V(\bx'),
\label{Born2}
\ee
where $\varphi$ is the angle between $\bq$ and $\bx'$.

\begin{figure}[htb]
  \begin{center}
    \leavevmode
\epsfxsize=7.5cm \epsffile{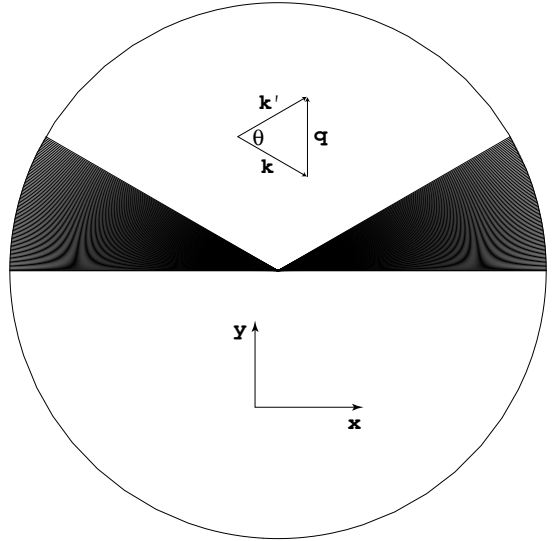}
  \end{center}
\caption{If the scattering is elastic, the DFT can be performed with $N$ shots. The 2D potential shown in Fig.~\ref{fig1} is located in the center of the ring. The length of the lines before and after the scattering are proportional to the length of the wavevectors $\bk$ and $\bk'$, respectively.} 
\label{fig2}
\end{figure}

For the implementation of the DFT we start first with a 2D potential
of the form 
$V(\bx')=V(r')(1+\cos \varphi )$.
If the scattering is elastic, we choose $\varphi=\pi/2-\theta/2$.
If the material allows inelastic scattering, we choose
$\varphi=\pi/2-\theta$. 
It will turn out that for this potential $\bq$ has to point always into the
same direction. For example $\bq$ can be always perpendicular
to the x-axis. Then we can define $\alpha$ 
as the angle between the x-axis and $\bx'$,
which turns our potential into $V(\bx')=V(r')(1+\sin \alpha )$.  
This 2D potential 
is shown in Fig.~\ref{fig1}. Inserting this potential into Eq. (\ref{Born2}) leads 
to
\be
f^{(1)} (\bk',\bk)=-\frac{\sqrt{2\pi } }{4} \frac{2m}{\hbar ^{2} } \int
r'dr'\left[ J_{0} (qr')+iJ_{1} (qr')\right] V(r'),
\label{Bessel}
\ee
which in the asymptotic limit 
$r'\gg 1/q$
becomes
\be
f^{(1)} (\bk',\bk)=-\frac{1}{2\sqrt{q} } \frac{2m}{\hbar ^{2} } \int
\sqrt{r'} dr'e^{i(qr'-\frac{\pi }{4} )} V(r').
\ee
By choosing 
$V(r')=\sum_{j=1}^Na_{j} \delta (r'-r_{j} )/\sqrt{r'}  $
we can encode the information into the $N$ amplitudes $a_{j} $
and perform a DFT
\be
f^{(1)} (\bk',\bk)=-\frac{1}{2\sqrt{q_{\nu}}} \frac{2m}{\hbar^{2} }
e^{-i\frac{\pi }{4} } \sum\limits_{j=1}^Na_{j} e^{iq_{\nu}r_{j} } 
\label{DFT}
\ee
In order for it to be a {\it discrete} Fourier transformation (DFT), 
we have to discretize the variables. Let us make a first attempt: 
$r_{j} =j\frac{l}{N} ,\quad j=1,2,\ldots ,N$,
and 
$q_{\nu } =\nu \frac{2\pi }{l} ,\quad \nu =1,2,\ldots ,N$, 
with $\nu$ referring to the $\nu$th detector. The smallest optically 
resolvable distance between the radii $r_j$ is $\lambda/2$. 
So it is safe to choose 
$l=N\lambda$, where $\lambda=2\pi/k$ is the deBroglie wavelength of the coherent beam. Then we 
obtain 
$\frac{\nu }{N} =2\sin \frac{\theta }{2} $
for the elastic scattering and
$\frac{\nu }{N} =\tan\theta$ for the inelastic scattering (see below).
The radii of the rings are multiples of the wavelength $\lambda$,
i.e.
$r_{j} =j\lambda $. There is still the requirement $r_{j}\gg 1/q_{\nu } $ that has to be satisfied.
Therefore we have to choose 
$j=N+1,N+2,\ldots ,2N$, which still produces a DFT, since the DFT is $2\pi$-periodic.

\begin{figure}[htb]
  \begin{center}
    \leavevmode
\epsfxsize=7.5cm \epsffile{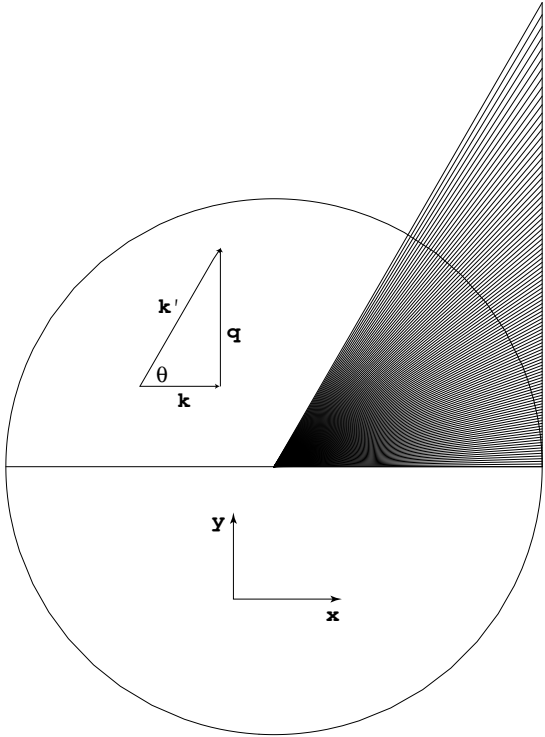}
  \end{center}
\caption{The 2D scattering potential shown in Fig.~\ref{fig1} is located at the center of the ring. The length of the lines before and after the scattering are proportional to the length of the wavevectors $\bk$ and $\bk'$, respectively. The DFT can be performed with a single shot if the scattering off the 2D scattering potential is inelastic. } 
\label{fig3}
\end{figure}

Our 2D potential $V(\bx')$ has the $1+\cos\varphi$ dependence, so that
we obtain $J_0$ and $J_1$ Besselfunctions in Eq.~(\ref{Bessel}). 
Since we do not want
to adjust the orientation of $V(\bx')$ according to the
scattering vector $\bq$, we have to keep the 
direction of $\bq$ constant in space. This requirement can be 
satisfied in two ways, which can be identified by elastic or 
inelastic scattering. In the case of elastic scattering $\bk$ 
and $\bk'$ have the same length. So 
$q:=\left| \bk-\bk'\right| =2k\sin \frac{\theta }{2} $, 
where $\theta$ is the angle between $\bk$ and $\bk'$. For 
every incoming beam of angle 
$\vartheta_{\rm in} =\theta /2$
the resulting vector element after the DFT can be read out in 
the detector at the outgoing angle 
$\vartheta _{\rm out} =\vartheta _{\rm in} =\theta /2$. 
The number of incoming and outgoing angles is exactly $N$. 
Thus the DFT can be performed with $N$ shots, which is shown 
in Fig.~\ref{fig2}. 
In the case of inelastic scattering, $\bk$ and $\bk'$ do not have 
the same length anymore. It is then possible to perform the DFT 
in a single shot. For this 
$q:=\left| \bk-\bk'\right| =k\tan \theta $. Since 
$k'=n_{r} k$
and 
$k=k'\cos \theta $, we obtain an angle-dependent refractive index 
$n_{r} =1/\cos \theta $, which is probably difficult to realize experimentally. 
At least in the case of a laser beam (see below) it may be possible to use 
photonic crystals that mimick $n_{r} =1/\cos \theta $. 
The inelastic scattering is shown in Fig.~\ref{fig3}.

\begin{figure}[htb]
  \begin{center}
    \leavevmode
\epsfxsize=8.5cm \epsffile{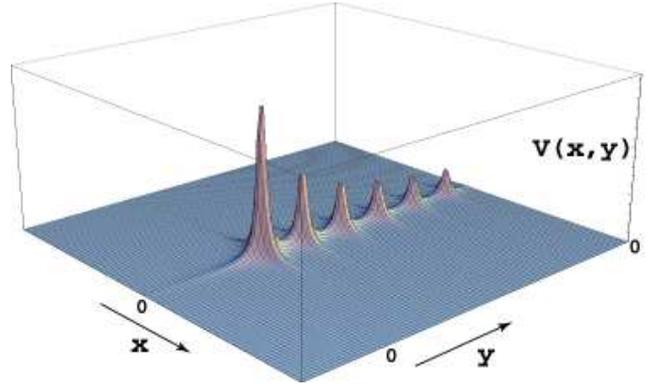}
  \end{center}
\caption{2D scattering potential $V(\bx')=\sum_{j=1}^Na_j\delta(r'-r_j)\delta(\alpha-\alpha_j)/r'$, which performs the DFT in a single shot for elastic scattering.} 
\label{fig4}
\end{figure}

Instead of 
$V(\bx')=(1+\sin \alpha )\sum\nolimits_{j=1}^Na_{j} \delta (r'-r_{j} )/\sqrt{r'}
$
consisting of concentric rings, 
we can alternatively use peaks of the form 
$V(\bx')=\sum_{j=1}^Na_{j} \delta (r'-r_{j} )\delta (\alpha -\alpha
_{j} )/r' $, which is shown in Fig.~\ref{fig4}. After putting the peaks on a straight 
line ($\alpha_{1}=\alpha_{2}=\dots =\alpha_N=\pi/2$), 
we obtain from Eq. (\ref{Born2})
\be
f^{(1)} (\bk',\bk)=-\frac{1}{4\sqrt{2\pi } } \frac{2m}{\hbar ^{2} }
\sum\limits_{j=1}^Na_{j} e^{iq_{\nu } r_{j} \cos \varphi _{\nu } },
\ee
where $\alpha_\nu=\varphi_\nu+\pi/2$ is the angle between
the scattering vector $\bq_\nu$ and the x-axis. 
With this potential we can discretize the variables similarly as before: 
$r_{j} =j\frac{l}{N} ,\quad j=1,2,\ldots ,N$,
and 
$q_{\nu } \cos \varphi _{\nu } =\nu \frac{2\pi }{l} ,\quad \nu =1,2,\ldots
,N$. The advantage of this method is that the peaks can be much 
easier produced experimentally than the rings. Also, we can start 
from the peak with $j=1$, because we do not require 
$r_{j}\gg 1/q_{\nu} $
anymore. There is also no scaling factor 
$1/\sqrt{q_{\nu}} $
needed for the detectors (see Eq. (\ref{DFT}))
and the scattering can be done
elastically with a single shot (see Fig.~\ref{fig5})
because the scattering vectors $\bq_{\nu}$ do not need to point
into the same direction.
However, the potential 
made of peaks decreases as 
$1/r'$, and not as 
$1/\sqrt{r'} $
as for the rings.

For the experimental realization of the 2D scattering off a 2D potential
one could use a 2D electron gas that is scattered off an electrostatic potential.
Like in the case of the laser beam scattering (see next section),
it would be also possible to use 3D scattering off a 2D potential,
such as the scattering of spin-polarized neutrons off a spatial 
electronic spin distribution.

\section{Laser beam scattering}
\label{lbs}

Since coherent matter waves have a very short deBroglie wavelength, 
the resolution of the 2D potential must be very high ($<1$ {\AA}), 
which is beyond today's technology. In order to overcome 
this difficulty, one can use 3D scattering of a laser beam off 
a 2D potential, where the wavelength of the laser can be chosen 
according to the accuracy of the 2D potential. For this we have 
to start from Maxwell's equations. Following Ref. \onlinecite{Jackson} we need to 
solve the Helmholtz equation
\bea
(\bnabla ^{2} +k^{2} )\bD & = & -\bnabla \times \bnabla \times (\bD-\epsilon_{0}
\bE) \nn\\
& & -\frac{i\epsilon _{0} \omega }{c} \bnabla \times (\bB-\mu _{0}\bH),
\eea
where $\bD$ and $\bB$ are the electrostatic and magnetic fields 
within the sample, $\bE$ and $\bH$ are the respective fields of 
the laser beam, $\epsilon_{0}$ is the dielectric constant, $\mu_{0}$ is 
the permeability, and 
$k=\sqrt{\epsilon_{0} \mu_{0}} \omega/c $
is the wavevector of our laser beam. The result reads\cite{Jackson}
\be
\bD=\bD^{(0)} +\bA_{s} e^{ikr}/r,
\ee
where the scattering amplitude in Born approximation 
is given by
\bea
\frac{\bepsilon^{\ast } \cdot \bA_{s}^{(1)} }{D_{0} } & = & \frac{k^{2} }{4\pi } \int
d^{3} x'\,e^{i\bq\cdot \bx'}\left\{\bepsilon^{\ast }\cdot\bepsilon_{0}\frac{\delta\epsilon
(\bx')}{\epsilon_{0}}\right. \nn\\
& & +\left.(\bn\times\bepsilon^{\ast } )\cdot(\bn_{0}\times\bepsilon_{0})
\frac{\delta\mu(\bx')}{\mu_{0}} \right\},
\eea
where $\delta\epsilon(\bx')\ll\epsilon_{0}$ and $\delta\mu(\bx')\ll\mu_{0}$.
Here $\bn_{0}$ and $\bepsilon_{0}$ are the direction and the 
polarization of the incoming beam, $\bn$ and $\bepsilon$ are 
the direction and the polarization of the outgoing beam, respectively. 
Let us restrict ourselves to variations in the dielectric constant 
$\delta \epsilon (x')$, i.e. 
$\delta \mu (x')=0$. 
In analogy to the 2D potential for the coherent matter beam, 
we can use a dielectric variation of the ring form 
$\delta \epsilon (\bx')=(1+\sin\alpha)\sum_{j=1}^Na_j\delta(r'-r_j)/\sqrt{r'}$
or of the peak form
$\delta \epsilon (\bx')=\sum_{j=1}^Na_{j} \delta (r'-r_{j} )\delta
(\alpha -\alpha _{j} )/r' $. Performing the 
integration in cylindrical coordinates, 
$\int d^{3} x'\, =\int r'dr'd\varphi \,dz $, for the peak form leads to
\be
\frac{\bepsilon^{\ast } \cdot \bA_{s}^{(1)} }{D_{0} } =\frac{k^{2} d}{4\pi \epsilon
_{0} } \bepsilon^{\ast } \cdot \bepsilon_{0} \sum\limits_{j=1}^Na_{j} e^{iq_{\nu } r_{j} \cos
\varphi _{\nu } },
\label{EM_scattering}
\ee
where $d$ is the thickness of the sample. So 3D scattering 
off a 2D potential provides another means to perform the DFT.

For the experimental implementation the spatial dielectric variation
$\delta \epsilon (\bx')$ can be produced for example electrostatically
by means of the Kerr effect.
As another example one could use a crystal with spin-orbit interaction
where an imbalance of spin-up and spin-down electrons leads
to different dielectric constants for left and right circularly
polarized light\cite{Kikkawa2000}. In this case a circularly polarized laser beam
is scattered off a spatial dielectric variation
$\delta \epsilon (\bx')$ that is created by a spatial electronic spin distribution.
It is also worth to mention that the laser scattering can be done at room temperature.

Now we estimate the amount of sampling points (=$N$) that can 
be used with today's technology. The state-of-the-art lasers of e.g. $\lambda=500$ 
nm wavelength produce an intensity of J=1 GW/cm$^{2}$ at the sample. 
This implies a photon number density of 
$n=\frac{J}{c\hbar \omega }\sim 10^{10}$ photons per (100 $\mu$m)$^{3}$. 
If only 10$^{-2}$ 
of the laser beam is scattered, about 10$^{4}$ photons remain at 
the detectors surrounding the sample at $r=1$ mm. Within a 
radius of 100 $\mu$m$=2N\lambda$ we can insert $N=100$ 
rings or peaks, i.e. use a basis 
of $N=100$ sampling points.

\begin{figure}[htb]
  \begin{center}
    \leavevmode
\epsfxsize=7.5cm \epsffile{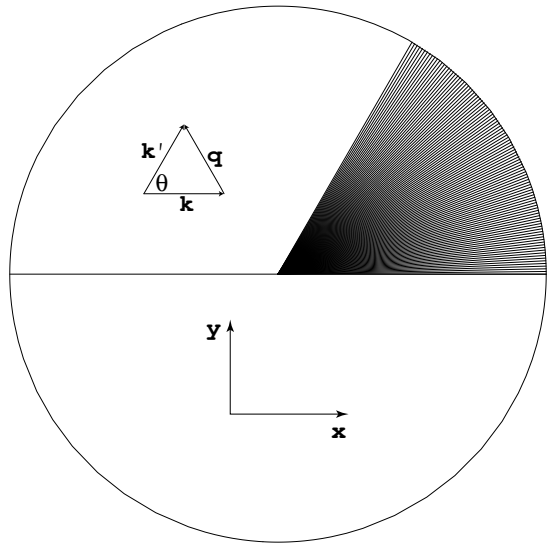}
  \end{center}
\caption{The 2D scattering potential shown in Fig.~\ref{fig4} is located in the center of the ring. The length of the lines before and after the scattering are proportional to the length of the wavevectors $\bk$ and $\bk'$, respectively. The DFT can be performed with a single shot by using elastic scattering off the 2D scattering potential. } 
\label{fig5}
\end{figure}

Until now we have taken a 2D variation of the dielectric constant 
$\delta \epsilon (\bx')=\sum_{j=1}^Na_{j} \delta (r'-r_{j} )\delta
(\alpha -\alpha _{j} )/r' $. 
For experiments it is more realistic to assume a 2D variation 
of the dielectric constant 
$\delta \epsilon (\bx')=\sum_{j=1}^Na_{j} \delta ^{(w)} (r'-r_{j}
)\delta ^{(w)} (\alpha -\alpha _{j} )/wr' $
made of delta functions of width $w$, such as 
$\delta ^{(w)} (r'-r_{j})=w/\pi [(r'-r_j)^{2}+w^{2}]$. 
Then Eq. (\ref{EM_scattering}) becomes approximately
\bea
\frac{\bepsilon^{\ast } \cdot \bA_{s}^{(1)} }{D_{0} } & = & 
\frac{k^{2}d}{4\pi \epsilon _{0} } \bepsilon^{\ast } \cdot \bepsilon_{0} 
\sum\limits_{j=1}^Na_{j}\int dr''
\frac{we^{iq_{\nu }(r''+r_j)\cos\varphi_{\nu}}}{\pi(r''^2+w^2)} \nn\\
& = & \frac{k^{2}de^{-q_{\nu}|\cos\varphi_{\nu}|w}}{4\pi \epsilon _{0} } 
\bepsilon^{\ast } \cdot \bepsilon_{0} 
\sum\limits_{j=1}^Na_{j}
e^{iq_{\nu } r_{j} \cos \varphi _{\nu } } ,
\label{EM_scattering_approx}
\eea
where we have made the substitution $r''=r'-r_j$.
As long as 
$w\ll 1/2k=\lambda/4\pi\leq 1/q_{\nu}|\cos\varphi_{\nu}|$, 
this exponential damping factor is negligible. Using 
a laser of wavelength $\lambda=500$ nm implies that 
$w\ll 50$ nm. It is also possible to scale up the proposed setup to longer 
wavelengths without changing the results. One could say that 
the required accuracy of the 2D potential is given by the wavelength 
of the laser. Therefore our scheme is very flexible and very 
fault-tolerant.\\
Since 
$\frac{\nu }{N} =2\sin \frac{\theta }{2} $ for elastic scattering, 
the detectors are to be arranged within a scattering angle 
of $\pi/3$. The smallest angle between two detectors is $\theta_{\rm min}=\pi/3N=\pi/300$, 
which leads to a minimum distance of $r\theta_{\rm min}=10$ $\mu$m 
between the detectors.

\section{Conclusion}

In conclusion, we have shown that by means of scattering a coherent 
matter or laser beam off a 2D potential it is possible to perform 
the DFT on $N$ sampling points. So the scattering
is done quantum-mechanically or classically.
The 2D potential can be made of concentric rings or 
peaks on a straight line.
The scattering off the peaks should be much easier
to implement experimentally since
it can be done elastically with a single shot.
Note that the DFT is performed classically in both cases
since we use a classical 2D potential to encode 
classical information.
This provides the great advantage that our device
can already be used as DFT coprocessor for existing classical
computers, without having to worry about the decoherence
of quantum states,
like in the case of quantum computing or spintronics 
devices\cite{Awschalom,Loss,Wolf,Kikkawa2000,Flatte}.
In addition, $M/N$ devices can be linked
together to perform the DFT in parallel on
$M$ sampling points, thereby making use of the FFT.
It should be possible with today's technology
to build a DFT device with the capability
to Fourier transform an $N=1024$-dimensional
vector.
If an
existing classical computer was equipped 
with $M/N=1024$ such DFT devices that are linked together, 
a speedup of a factor $M=1024N\sim 10^{6}$ could
be achieved. 

\section{Acknowledgments}

This work has been supported in part by the Swiss NSF, NCCR Nanoscience, 
U.S. NSF, and DARPA/ARO. 
We would like to thank Ryan Epstein and Martino Poggio for useful discussions.

\end{document}